\documentclass[12pt]{article}
\usepackage{graphicx}
\usepackage{cite}

\def \bea{\begin{eqnarray}}
\def \beq{\begin{equation}}
\def \b{{\cal B}}

\def \eea{\end{eqnarray}}
\def \eeq{\end{equation}}

\def \s{\sqrt{2}}

\def \3half{\frac{3}{2}}

\begin{document}
\begin{flushright}
EFI 08-19 \\
June 2008 \\
\end{flushright}
\centerline{\bf $B$ decays dominated by $\omega$--$\phi$ mixing}
\bigskip
\centerline{Michael Gronau\footnote{On sabbatical leave from the Physics 
Department, Technion, Haifa 32000, Israel.} and Jonathan L. Rosner}
\medskip
\centerline{\it Enrico Fermi Institute and Department of Physics,
 University of Chicago} 
\centerline{\it Chicago, IL 60637, U.S.A.} 
\bigskip
\begin{quote}
Recently Belle has established the 90\% confidence level (CL) upper
limit $\b < 9.4 \times 10^{-7}$ for the branching ratio for $B^0\to J/\psi
\phi$, a process expected to be suppressed by the Okubo-Zweig-Iizuka (OZI)
rule disfavoring disconnected quark diagrams.
We use information on
$\omega$--$\phi$ mixing to establish likely lower bounds on this and related
processes.  We find that the Belle result is about a factor of five above
our limit, while other decays such as $B^0 \to \bar D^0 \phi$ and $B^+
\to \pi^+ \phi$, for which upper limits have been obtained by BaBar,  could be
observable with similar improvements in data.
We argue that a significant enhancement of our predicted decay rates by
rescattering is unlikely.
\end{quote} 

\leftline{\qquad PACS codes:  12.15.Hh, 12.15.Ji, 13.25.Hw, 14.40.Nd} 

\medskip
Certain $B$ meson decay processes are expected to be suppressed, as they
involve either the participation of a spectator quark or, equivalently in
terms of flavor topology, the rescattering of final states into one another.
A number of such processes were suggested in Ref.\ \cite{Blok:1997yj}.
Recently the Belle Collaboration has greatly improved upper limits on one
such process, the decay $B^0 \to J/\psi \phi$, finding a 90\% CL upper limit
for the branching ratio of $\b < 9.4 \times 10^{-7}$ \cite{Liu:2008bt}. 
This process is expected to be suppressed by the Okubo-Zweig-Iizuka (OZI)
rule disfavoring disconnected quark diagrams~\cite{Okubo:1963fa}.
In the present Letter we evaluate one potential mechanism for production of
such a final state, the existence of $\omega$--$\phi$ mixing, and find that
it leads to a predicted branching ratio about a factor of five below the
present upper limit.  We also evaluate the effect of this mixing upon
several other processes and find that for $B^0 \to \bar D^0 \phi$ and $B^+
\to \pi^+ \phi$ a similar improvement in data should lead to observable
signals.

Extensive discussions of $\omega$--$\phi$ mixing have been given in Refs.\
\cite{Benayoun:1999fv,Benayoun:1999au,Benayoun:2000ti,Benayoun:2001qz,%
Benayoun:2007cu}.  We shall neglect isospin-violation and admixtures with
the $\rho$.  Then one can parametrize $\omega$--$\phi$ mixing in terms of an
angle $\delta$ such that the physical $\omega$ and $\phi$ are related to the
ideally mixed states $\omega^I \equiv (u \bar u + d \bar d)/\s$ and $\phi^I
\equiv s \bar s$ by
\beq\label{mixing}
\left( \begin{array}{c} \omega \\ \phi \end{array} \right) = 
\left( \begin{array}{c c}
\cos \delta & \sin \delta \\ - \sin \delta & \cos \delta
\end{array} \right)
\left( \begin{array}{c} \omega^I \\ \phi^I \end{array} \right)
\eeq
A simplified analysis \cite{Benayoun:1999fv} implies a mixing angle
of $\delta = - (3.34 \pm 0.17)^\circ$, while the most 
recent treatment~\cite{Benayoun:2007cu} implies an energy-dependent mixing 
which varies from $-0.45^\circ$ at the $\omega$ mass to $-4.64^\circ$ at 
the $\phi$ mass.

\begin{figure}
\centerline{\includegraphics[width=6.0cm]{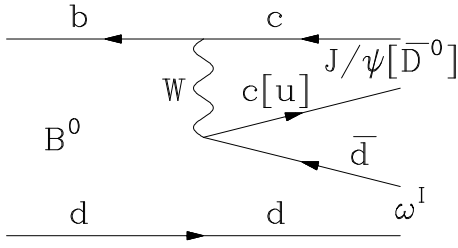}}
\caption{Quark diagrams for OZI-allowed $B^0\to J/\psi \omega^I$ 
and $B^0\to \bar D^0\omega^I$.}
\label{fig:csjpsi}
\end{figure}
%
\begin{figure}
\centerline{\includegraphics[width=5.0cm]{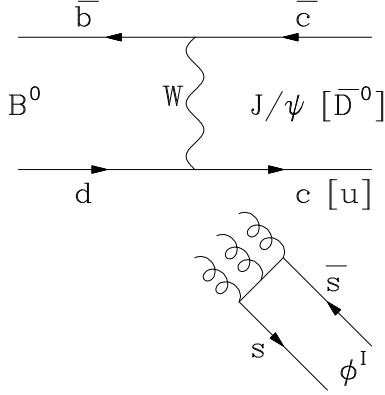}}
\caption{Quark diagrams for OZI-suppressed $B^0\to J/\psi \phi^I$ and 
$B^0\to \bar D^0\phi^I$.}
\label{fig:jpsiphi}
\end{figure}

A systematic study of  the effects of $\omega$--$\phi$ 
mixing on hadronic decays of non-strange $B$ mesons, $B^0\equiv \bar bd, 
B^+\equiv \bar bu$,  requires considering the three $\Delta S=0$ quark 
subprocesses, $\bar b \to \bar cu\bar d, \bar b\to \bar c c\bar d$ 
and $\bar b\to \bar u u\bar d$. Each one of these subprocesses leads
to OZI-allowed decays involving $\omega^I$, while decays into final states
with $\phi^I$ are OZI-suppressed. 
Quark diagrams describing two examples of OZI-allowed decays, $B^0\to J/\psi \omega^I$
and $B^0\to \bar D^0\omega^I$, and  corresponding OZI-suppressed decays, 
$B^0\to J/\psi\phi^I$ and $B^0\to \bar D^0\phi^I$, are shown in Figs.~\ref{fig:csjpsi}  and
\ref{fig:jpsiphi}. The first two processes are described by color-suppressed tree diagrams, 
while the other two processes involve $W$-exchange diagrams, to which an $s\bar s$ pair
is attached through three gluons.

\begin{figure}
\centerline{\includegraphics[width=6.0cm]{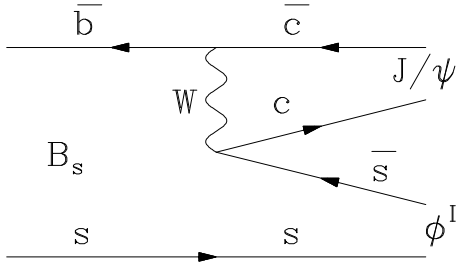}}
\caption{Quark diagrams for OZI-allowed $B_s\to J/\psi \phi^I$.}
\label{fig:csbs}
\end{figure}

\begin{figure}
\centerline{\includegraphics[width=5.0cm]{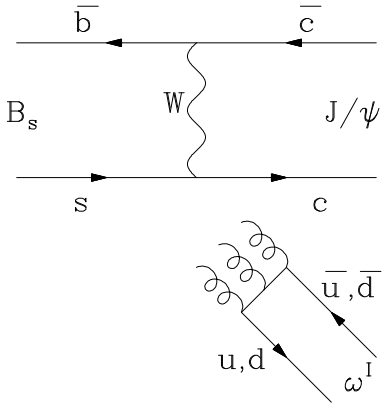}}
\caption{Quark diagrams for OZI-suppressed $B_s\to J/\psi \omega^I$.}
\label{fig:bsjpsiom}
\end{figure}

The situation in decays of $B_s\equiv \bar bs$ is the opposite relative to that
in non-strange $B$ decays. That is, the $\omega$ and $\phi$ exchange roles. 
Here one considers the $\Delta S=1$ quark subprocess $\bar b\to \bar c c\bar s$
which leads to OZI-allowed decays involving $\phi^I$ and OZI-suppressed decays
with $\omega^I$.  (The quark subprocess $\bar b\to \bar c\bar us$ leads through
W-exchange diagrams to OZI-allowed $B_s$ decays involving $\omega^I$ including
$\bar D^0\omega^I$.) Examples of quark diagrams describing the OZI-allowed
decay $B_s\to J/\psi \phi^I$ and the corresponding OZI-suppressed decay $B_s\to
J/\psi \omega^I$ are shown in Figs.~\ref{fig:csbs} and \ref{fig:bsjpsiom}.  As
in the above examples of $B^0$ decays, the first process is governed by a
color-suppressed tree amplitude, while the second decay is described by a
W-exchange diagram, to which a $u\bar u$ or $d\bar d$ pair is attached by three
gluons.

Neglecting contributions of OZI-suppressed amplitudes and small phase space 
differences between processes with $\omega$ or $\phi$ in the final state,
Eq.~(\ref{mixing}) implies
\bea\label{Xphi}
\b(B^{0,+} \to X^{0,+}\phi) &=& \tan^2\delta\,\b(B^{+,0}\to X^{+,0}\omega)~,\\
\label{Xomega}
\b(B_s \to X^0\omega) &=& \tan^2\delta\,\b(B_s\to X^0\phi)~.
\eea
The examples shown in Figs.~\ref{fig:csjpsi} and \ref{fig:csbs} correspond to 
$X^0=J/\psi, \bar D^0$ in $B^0$ decays and $X^0=J/\psi$ in $B_s$ decays.

%

In Table I we list OZI-allowed branching ratios of $B^0, B^+$ and 
$B_s$ decays, for which  nonzero values have been measured, and upper limits on
corresponding OZI-suppressed decays. 
The upper left part of the table, addressing non-strange $B$ 
mesons, includes also processes involving $\rho^0$, for which
branching ratios are expected to be approximately equal to corresponding 
processes with $\omega$.  (See Fig.~\ref{fig:csjpsi}.)
The approximately equal decay rates measured for 
$B^0\to\bar D^0\rho^0$ and $B^0\to \bar D^0\omega$  confirm this assumption. 
OZI-allowed branching ratios for $B_s$ decays involving $\phi$ are listed   
in the lower left part of the Table I. 

Using Eqs.~(\ref{Xphi}) and (\ref{Xomega}) with a universal value of $\delta = -4.64^\circ$ 
and the measured OZI-allowed branching ratios, we obtain predictions for the 
OZI-suppressed rates shown in the right-hand column of Table I.  
Values in parentheses, quoting predictions for OZI-suppressed $B_s$ decays 
involving $\omega$,  are obtained for the small mixing angle 
$\delta = -0.45^\circ$~\cite{Benayoun:2007cu}.
Predictions for $B^0$ and $B^+$ decays are compared with current upper bounds measured for 
these branching ratios. We note that the predictions for $\b(B^0\to \bar D^0\phi)$, 
$\b(B^0\to J/\psi\phi)$ and $\b(B^+\to \pi^+\phi)$ are about a factor five smaller
than the current upper limits on these branching ratios.

\renewcommand{\arraystretch}{1.1}
\begin{table}
\caption{Comparison of some OZI--allowed and OZI--suppressed branching ratios,
in units of $10^{-5}$ and $10^{-7}$, respectively. Averages are taken from 
Ref.~\cite{PDGup}.  Upper limits 
are 90\% CL.  Predictions are based on an $\omega$-$\phi$ mixing angle
$\delta= -4.64^\circ$. Parentheses denote predictions based on the very small
admixture of $s \bar s$ expected in the $\omega$ in Ref.\
\cite{Benayoun:2007cu}.
\label{tab:comp}}
\begin{center}
\begin{tabular}{r r c r c c} \hline \hline
\multicolumn{1}{c}{~} & \multicolumn{2}{c}{OZI--allowed}
 & \multicolumn{3}{c}{OZI--suppressed} \\
Quark~~~~& Decay & ${\cal B}~(10^{-5})$ & Decay
 & \multicolumn{2}{c}{${\cal B}~(10^{-7})$} \\
subprocess & mode  & ~ & mode & Upper limit & Predicted\\ \hline
 $\bar b\to\bar cu\bar d$ & $B^0 \to \bar D^0 \rho^0$ 
 & 32$\pm$5~\cite{Kuzmin:2006mw} & 
 $\bar D^0 \phi$  &   $< 116$~\cite{Aubert:2007nw}  & 21$\pm$3 \\
 ~  & $\bar D^0 \omega$ & 25.9$\pm$3.0~\cite{Aubert:2003sw} & $\bar D^0 \phi$ 
 & $< 116$~\cite{Aubert:2007nw}  & 17$\pm$2\\
 ~ & $\bar D^{*0} \omega$ & 27$\pm$8~~\cite{Aubert:2003sw} & $\bar D^{*0} \phi$ 
 & -- & 18$\pm$5\\
$\bar b \to \bar cc\bar d$ & $J/\psi \rho^0$ & 2.7$\pm$0.4~\cite{Aubert:2007xw}
 & $J/\psi \phi$ & $< 9.4$~\cite{Liu:2008bt} & 1.8$\pm$0.3 \\
$\bar b\to \bar uu\bar d$ & $B^+ \to\pi^+\omega$ & 0.69$\pm$0.05~\cite{Jen:2006in}
 & $B^+ \to \pi^+\phi$ & $<2.4$~\cite{Aubert:2006nn} & 0.45$\pm$0.03 \\
  ~ & $\rho^+ \omega$ & $1.06^{+0.26}_{-0.23}$~\cite{Aubert:2006vt} 
  & $\rho^+ \phi$ & $<160$~\cite{Bergfeld:1998ik} & 0.7$\pm$0.2 \\ \hline
$\bar b\to \bar cc\bar s$ & $B_s \to J/\psi \phi$ & 93$\pm$33~\cite{Abe:1996kc}
 & $B_s \to J/\psi \omega$ & -- & 61$\pm$22~(0.6) \\
~ & $\psi(2S) \phi$ & 48$\pm$22~\cite{Abulencia:2006jp} & $\psi(2S) \omega$ 
& -- & 32$\pm 15$~(0.3)\\
\hline \hline
\end{tabular}
\end{center}
\end{table}

\begin{figure}[b]
\centerline{\includegraphics[width=6.0cm]{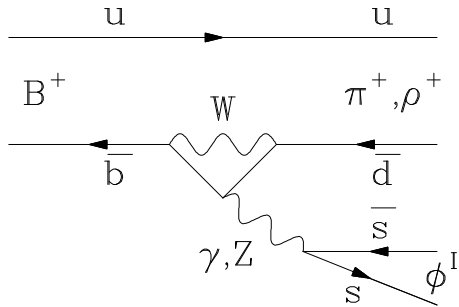}}
\caption{Electroweak penguin diagram for OZI-suppressed $B^+\to \pi^+\phi^I$
and $B^+\to \rho^+\phi^I$.}
\label{fig:ewp}
\end{figure}

\begin{figure}
\centerline{\includegraphics[width=6.0cm]{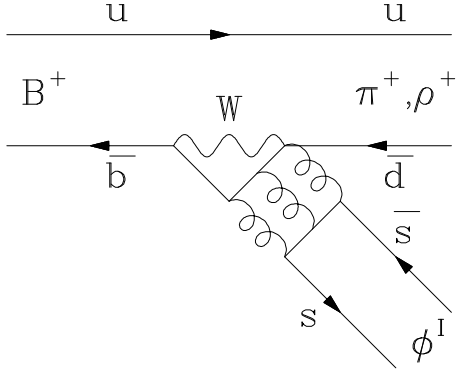}}
\caption{Singlet penguin diagram for OZI-suppressed $B^+\to \pi^+\phi^I$
and $B^+\to \rho^+\phi^I$.}
\label{fig:p3g}
\end{figure}

While most OZI-allowed processes in Table I are described by color-suppressed
tree diagrams as shown in Figs.~\ref{fig:csjpsi} and \ref{fig:csbs},
the CKM-suppressed charmless decays $B^+\to \pi^+\omega$ and $B^+\to
\rho^+\omega$ are dominated by color-allowed tree
diagrams~\cite{Gronau:1994rj,Chiang:2003pm}. Contributions to these processes
from color-suppressed tree diagrams are considerably smaller. 
This is demonstrated by 90$\%$ CL upper limits measured for corresponding 
color-suppressed branching ratios,
$\b(B^0\to \pi^0\omega) < 0.12\times 10^{-5}$~\cite{Aubert:2004ih}
and $\b(B^0\to \rho^0\omega) < 0.15\times 10^{-5}$~\cite{Aubert:2006vt}, which are
a factor six or seven below $\b(B^+\to \pi^+\omega)$ and 
$\b(B^+\to \rho^+\omega)$ given in Table I.
The corresponding OZI-suppressed amplitudes for $B^+\to \pi^+\phi$ and 
$B^+\to \rho^+\phi$ each obtain an electroweak penguin 
contribution~\cite{Gronau:1995hn} and a contribution 
from a singlet penguin diagram~\cite{Dighe:1997wj}, shown in Figs.~\ref{fig:ewp}
and~\ref{fig:p3g}, respectively.
These amplitudes have been calculated in Ref.~\cite{Beneke:2003zv} 
and~\cite{Beneke:2006hg} within the framework of QCD factorization neglecting
$\omega$-$\phi$ mixing. Branching ratios $\b(B^+\to \pi^+\phi)=(2-10)\times
10^{-9}$ and $\b(B^+\to \rho^+\phi)=(1-3)\times 10^{-8}$ were obtained,
considerably smaller than the two corresponding predictions in Table I
originating in $\omega$-$\phi$ mixing. 

Assuming that small OZI-suppressed amplitudes do not 
interfere destructively with amplitudes due to $\omega$-$\phi$ mixing, 
the predictions presented in Table I for branching ratios of OZI-suppressed 
decays should be considered as 
likely lower bounds. In principle, these branching ratios 
may be enhanced by rescattering through intermediate states with larger
decay rates. This possibility had been envisaged a few years before starting
the operation of $e^+e^-$ $B$ factories~\cite{Blok:1997yj}.  We will now argue 
that experimental evidence obtained in certain experiments 
indicates that a significant enhancement by rescattering is unlikely in OZI-suppressed 
and other suppressed decays.

Consider the decay $B^0\to D^-_sK^+$, which is governed by a $W$-exchange 
amplitude represented by a quark subprocess $(\bar bd)\to (\bar cu)$,
associated with a popping of an $s\bar s$ pair out of the vacuum. This exchange
amplitude is expected to be suppressed by an order of magnitude ($\sim
\Lambda_{\rm QCD}/m_b$) relative to the corresponding color-favored tree
amplitude for  $B^0\to D^-\pi^+$ induced by the same quark subprocess,
$\bar b\to \bar cu\bar d$~\cite{Gronau:1994rj,Beneke:2000ry,Mantry:2003uz}. 
This would imply $\b(B^0\to D^-_s K^+)/\b(B^0\to D^-\pi^+)\sim 10^{-2}$.
Rescattering through dynamically favored intermediate states including
 $B^0\to D^-\pi^+\to D^-_sK^+$ and rescattering through other intermediate $C=-1, S=0$ 
 states, with decay branching ratios at a level of a fraction of a percent, could enhance 
 the branching ratio for $B^0\to D^-_sK^+$ relative to the above expectation. 
 Experimentally, one 
finds~\cite{Krokovny:2002pe} $\b(B^0\to D^-_sK^+)=(2.9\pm 0.5)\times 10^{-5}$,
in comparison with~\cite{PDGup} $\b(B^0\to D^-\pi^+)=(2.68\pm 0.13)\times10^{-3}$ which 
is two orders of magnitude larger. That is, rescattering effects do not enhance 
the rate for $B^0\to D^-_sK^+$ beyond  the estimate based on an exchange 
amplitude. 

One possible conclusion is that a significant enhancement  of
diagramatically suppressed decay rates by rescattering  requires 
intermediate states with rates which are larger than the suppressed rates 
by {\em more than two orders of magnitude}. This requirement seems to follow from the
multi-channel nature of the rescattering process occurring between the initial
$B$ meson and the final state to which it decays.  Examples for processes which
have been shown to need an enhancement by rescattering are the decays
$B\to K\pi$~\cite{Ciuchini:1997rj}.  The short-distance loop-suppressed penguin
amplitude dominating these processes is too small to account for the measured
decay rates and requires an enhancement by long-distance
rescattering~\cite{Jain:2007dy}. The branching ratios of intermediate states
including  $B\to D^{(*)-}_s D^{(*)}$ are at a percent level, {\em three orders
of magnitude} larger than the branching ratios calculated for $B\to K\pi$ using
short short-distance physics.  This is sufficient for a significant enhancement
of the $B\to K\pi$ decay rates relative to this calculation.

A well-known charmless $B$ decay process dominated by a $W$-exchange 
amplitude is $B^0\to K^+K^-$~\cite{Blok:1997yj,Gronau:1994rj}. 
This process receives rescattering contributions from 
tree-dominated intermediate states including $\pi^+\pi^-$ 
with a branching ratio~\cite{PDGup} 
$\b(B^0\to\pi^+\pi^-) = (5.13\pm 0.24)\times 10^{-6}$. Assuming an order of 
magnitude suppression of the exchange amplitude relative to a tree amplitude 
as in $B^0\to D^-_sK^+$,
and using the above criterion for no significant enhancement by rescattering, 
one expects $\b(B^0\to K^+K^-) \sim 5\times 10^{-8}$, almost an order of magnitude
below the current 90\% CL upper limit of~\cite{PDGup} $4.1\times 10^{-7}$.
Similarly, using~\cite{PDGup} $\b(B^0\to\rho^+\rho^-) = (2.42\pm 0.31)\times 10^{-5}$, 
we predict  $\b(B^0\to K^{*+}K^{*-})\sim 2\times 10^{-7}$. Very recently an upper limit 
at 90\% CL has been measured~\cite{Aubert:2008ap}, 
$\b(B^0\to K^{*+}K^{*-}) <2.0\times 10^{-6}$, an order of magnitude above our prediction.

\begin{figure}
\centerline{\includegraphics[width=0.98\textwidth]{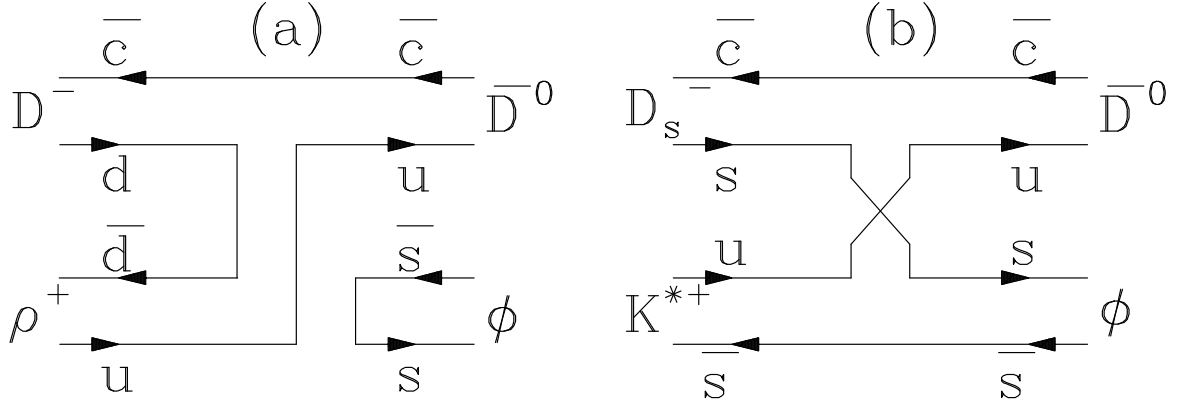}}
\caption{Quark diagrams for (a) OZI-suppressed rescattering $D^- \rho^+
\to \bar D^0 \phi$, (b) OZI-allowed rescattering $D_s^- K^{*+} \to \bar D^0
\phi$.
\label{fig:OZI}}
\end{figure}

Consider now the OZI-suppressed decay $B^0\to \bar D^0\phi$. The quark diagram 
for this process shown in Fig.~\ref{fig:jpsiphi} describes an exchange amplitude 
$(\bar bd)\to (\bar cu)$ as in $B^0\to D^-_sK^+$, to which a pair of $s\bar s$ is 
attached by three gluons. The above argument for no significant rescattering effects in 
$B^0\to D^-_sK^+$ implies the absence of such effects also in $B^0\to \bar D^0\phi$.
To demonstrate this explicitly, let us consider the two kinds of intermediate states through 
which rescattering into $\bar D^0\phi$ can occur: 
\begin{enumerate}
\item States dominated by tree amplitudes such as $D^-\rho^+~[(\bar cd)(\bar du)]$.  Rescattering 
through these intermediate states into $\bar D^0\phi$ $[(\bar cu)(\bar ss)]$ is OZI-suppressed 
[see Fig.~\ref{fig:OZI}(a)] and is not expected to enhance the rate for $B^0\to\bar D^0\phi$.
\item States governed by exchange amplitudes  including $D^-_sK^{*+}~[(\bar cs)(\bar su)]$. 
Rescattering  through these states into $\bar D^0\phi~[(\bar cu)(\bar ss)]$ is OZI-allowed [see 
Fig.~\ref{fig:OZI}(b)]. This rescattering 
is not expected to enhance the predicted branching ratio, $\b(B^0\to \bar D^0\phi)\simeq
2\times 10^{-6}$, because  $\b(B^0\to D^-_sK^{*+})$ is expected to be only 
one order of magnitude larger, assuming that it does not differ much from 
$\b(B^0\to D^-_sK^+)=(2.9\pm 0.5)\times 10^{-5}$ .
\end{enumerate}

A similar situation exists in $B^0\to J/\psi\phi$. Here OZI-suppressed
rescattering occurs through intermediate tree-dominated states including
$D^{(*)+}D^{(*)-}$, while OZI-allowed rescattering involves intermediate states
such as $D^{(*)+}_sD^{(*)-}_s$. $B^0$ decays into the latter states are
dominated by exchange amplitudes, which are expected to be suppressed by about
an order of magnitude relative to the tree amplitudes in $B^0\to
D^{(*)+}D^{(*)-}$.
Using the suppression measured for the exchange amplitude in $B^0\to D^-_sK^+$,
this implies, for instance~\cite{Gronau:2008ed}, $\b(B^0\to D^+_sD^-_s) =
(4.0^{+1.8}_{-1.4})\times 10^{-6}$, an order of magnitude below the current
upper limit on this branching ratio~\cite{Zupanc:2007pu}.  This branching ratio
is only twenty times larger than the value predicted for
$\b(B^0\to J/\psi\phi)$, which is expected to be insufficient for an
enhancement of the latter branching ratio by rescattering through this class of
intermediate states.

In the case of $B\to \pi^+\phi$ (or $B^+\to\rho^+\phi$) the situation is
slightly different, but the condition for a significant enhancement by
rescattering is also not met. The final state $\pi^+\phi$ may be reached by
OZI-suppressed rescattering through intermediate states such as $\pi^+\rho^0$,
or by OZI-allowed rescattering through states including $K^+\bar K^{*0}$. $B^+$
decay into the latter mode is dominated by a suppressed $\Delta S=0$ penguin
amplitude~\cite{Chiang:2003pm} implying a small branching ratio.  The current
90$\%$ CL upper limit~\cite{Aubert:2007ua}, $\b(B^+\to K^+\bar K^{*0})<
1.1\times 10^{-6}$, shows that this branching ratio is at most twenty-five
times larger than the predicted $\b(B^+\to\pi^+\phi)$. As discussed above,
this is insufficient for enhancing the latter branching ratio by rescattering.

In conclusion, we have studied the consequences of $\omega$--$\phi$ mixing in 
OZI-suppressed hadronic decays of $B$ and $B_s$ mesons. We calculated branching
ratios for $B$ decays involving $\phi$, which in the cases of $B^0\to \bar
D^0\phi$, $B^0\to J/\psi\phi$ and $B^+\to \pi^+\phi$ are each about a factor
of five below the corresponding current upper limits. We used the observed
suppression of branching ratios for decays dominated by $W$-exchange including
$B^0\to D^-_sK^+$ to argue that a significant enhancement of these rates by
rescattering is unlikely.  Thus, the above three processes are predicted to be 
detectable with a factor of five increase in data.
Effects of $\omega$--$\phi$ mixing in OZI-suppressed $B_s$ decays involving 
$\omega$ are much smaller than in nonstrange $B$ decays  if one assumes a very 
small admixture of $s \bar s$ 
in the $\omega$ as suggested in Ref.\ \cite{Benayoun:2007cu}.
The predicted branching ratios become a factor two smaller than in Table I for 
an energy-independent $\omega$-$\phi$ mixing angle of 
$\delta=-3.34^\circ$~\cite{Benayoun:1999fv}.
\bigskip

M.G. would like to thank the Enrico Fermi Institute at 
the University of Chicago for its kind and generous hospitality. We thank 
Pavel Krokovny, Shunzo Kumano, and Yoshi Sakai for 
useful communications.  This work was supported in part by 
the United States Department of Energy through Grant No.\ DE FG02 90ER40560.

\end{document}